\documentclass[doublecol]{epl2} 

\usepackage{amsmath}
\usepackage{amsfonts}
\usepackage{amssymb}
\usepackage{bbm}

\title{Emergence of gravity from spinfoams}
\shorttitle{Emergence of gravity from spinfoams} %Insert here a short version of the title if it exceeds 70 characters

\author{Elena Magliaro \and Claudio Perini}
\shortauthor{E. Magliaro and C. Perini}

\institute{                    
Institute for Gravitation and the Cosmos, Physics Department, Penn State, University Park, PA 16802-6300, USA
}

\abstract{We find a nontrivial regime of spinfoam quantum gravity that reproduces classical Einstein equations. This is the double scaling limit of small Immirzi parameter (gamma), large spins (j) with physical area (gamma times j) constant. In addition to quantum corrections in the Planck constant, we find new corrections in the Immirzi parameter due to the quantum discreteness of spacetime. The result is a strong evidence that the spinfoam covariant quantization of general relativity possesses the correct classical limit.
}

\begin{document}

\newcommand{\mc}[1]{\mathcal{#1}}
\newcommand{\mbb}[1]{\mathbbm{#1}}
\newcommand{\Tr}{{\rm Tr}}
\newcommand{\ket}[1]{|#1\rangle}
\newcommand{\varket}[1]{[#1\rangle}
\newcommand{\bra}[1]{\langle#1|}
\newcommand{\comment}[1]{}

\newcommand{\im}[1]{\text{Im}\,#1}
\newcommand{\re}[1]{\text{Re}\,#1}

\maketitle

\section{Introduction}
Spinfoams \cite{Baez:1997zt,Reisenberger:2000fy,Perez:2003vx} are a tentative covariant quantization of general relativity. They provide transition amplitudes between quantum states of 3-geometry, in the form of a Misner-Hawking sum over virtual geometries \cite{Misner:1957wq,Hawking:1979zw}. In the so-called `new models' \cite{Engle:2007wy,Livine:2007ya,Freidel:2007py}, intermediate quantum states are the ones of canonical loop quantum gravity, the $SU(2)$ spin-network states, a remarkable feature that promotes the spinfoam framework to a tentative path integral representation of loop quantum gravity.

The physical picture emerging from the spinfoam gravity is the following: spacetime is a quantum foam of virtual geometries with a discrete and purely combinatorial, relational structure, where the Planck scale plays the role of a natural minimal length. %Spinfoams are the result of the quantization of general relativity formulated as a constrained BF theory. In the BF theory with gauge group $G$, the basic variables are a 2-form $B$ and a connection 1-form $\omega$, both valued in the Lie algebra of $G$. The equations of motion of BF theory impose the flatness of the connection
%\begin{align}
%F(\omega)=d\omega+\omega\wedge\omega=0
%\end{align}
%where $F$ is the curvature. So this theory has no local degrees of freedom. 

General relativity is described by the Holst action
\begin{align}\label{Sintro}
S=\int{}^*(e\wedge e)\wedge F(\omega)+\frac{1}{\gamma}\int (e\wedge e)\wedge F(\omega)
\end{align}
with gauge group $SO(1,3)$ (or $SO(4)$ in the Euclidean signature), where the second term vanishes by using the equations of motion. The first term is the Einstein-Hilbert-Palatini action for general relativity in terms of the cotetrad $e$ and the connection $\omega$, regarded as independent variables ($F$ is the curvature). The real number $\gamma\neq0$ is called Barbero-Immirzi, or Immirzi parameter, and controls the spacetime discreteness as it enters the discrete spectra of area and volume operators \cite{Rovelli:1994ge,Ding:2010ye}. Spinfoam models provide a Feynman path integral, or state sum, based on a discretization of \eqref{Sintro} over a 2-complex (generalized triangulation) and the full degrees of freedom are recovered in the infinite refinement limit or equivalently summing over 2-complexes \cite{Rovelli:2010qx}. The spinfoam theory is sufficiently simple \cite{Rovelli:2010bf} and possesses the correct symmetries \cite{Rovelli:2010ed}. Recently it has been successfully coupled to matter fields \cite{Bianchi:2010bn,Han:2011as}. Furthermore, the large distance analysis was able to extract the correct low-energy physics in some simple cases \cite{Bianchi:2006uf,Alesci:2007tx,Bianchi:2009ri,Bianchi:2010zs}. 

A major open problem is to show that the spinfoam Feynman path integral is able to reproduce Einstein equations when a semiclassical expansion is performed. What we would like to have is the spinfoam version of the following semiclassical expansion of the gravitational path integral
\begin{align}\label{whatweexpect}
\int Dg_{\mu\nu}e^{\frac{i}{\hbar}S_{EH}(g_{\mu\nu})}\sim e^{\frac{i}{\hbar}S_{EH}(g^0_{\mu\nu})}
\end{align}
in the classical limit $\hbar\rightarrow 0$, where $S_{EH}$ is the Einstein-Hilbert action for general relativity and on the right hand side it is evaluated on the classical solution $g^0_{\mu\nu}$ of the equations of motion determined by the boundary conditions on the metric field $g_{\mu\nu}$. Here we propose a solution to this problem. 

In this letter we improve and generalize previous arguments \cite{Magliaro:2011qm} to a general 2-complex. In the continuous area spectrum limit $\gamma\rightarrow 0$ and in the semiclassical limit $\hbar\rightarrow 0$ we find the analogous of the WKB expansion \eqref{whatweexpect} for spinfoam quantum general relativity. We can state the regime in a more suggestive fashion by introducing two physical scales. One is the length scale of \emph{quantum gravity}, identified with the Planck length
\begin{align}
l_{QG}\equiv l_P,
\end{align}
the other is the scale of \emph{loop quantum geometry}, that is the scale where we can `see' the discreteness of spacetime
\begin{align}
l_{LQG}\equiv\sqrt{\gamma}\,l_P.
\end{align}
Thus the regime of the spinfoam path-integral we look for is expressed by the following relation:
\begin{align}
l\gg l_{QG}\gg l_{LQG}
\end{align}
where $l$ is the typical linear scale of each 4-simplex in the spacetime triangulation. The analysis is partly based on the path integral formulation of references \cite{Conrady:2008ea,Conrady:2008mk}. Throughout the letter we work in natural units $G=\hbar=c=1$, but will restore some of those constants when needed. In particular, restoring only $\hbar$ the Planck length is $l_P=\sqrt\hbar$. 
\section{The spinfoam amplitude}
We consider the spinfoam amplitude \cite{Engle:2007wy,Freidel:2007py} for a 2-complex $\sigma$ without matter. We restrict for simplicity to the Euclidean signature and to Barbero-Immirzi parameter $0<\gamma<1$, where formulas get simpler. For each face $f$ (the 2-cells) of the 2-complex $\sigma$ there is an associated integer spin $j_f$. Faces are oriented and bounded by a cycle of edges $e$ (the 1-cells). Each edge bounding a face has a source vertex $s(e)$ and a target vertex $t(e)$, where source/target is relative to the orientation of the face. To each edge let us associate $SU(2)$ elements $n_{ef}$ ($f$ runs over the faces meeting at the edge $e$) and two source/target $Spin(4)\simeq SU(2)\times SU(2)\sim SO(4)$ gauge group variables $g_{e,s(e)}$, $g_{e,t(e)}$. The variables $n_{ef}$ can also be interpreted as unit vectors $\vec n_{ef}$ in $\mathbbm R^3$, up to a phase ambiguity, by saying that $n$ is a rotation that brings a reference direction to the direction of $\vec n$.

The spinfoam amplitude, or partition function, for the 2-complex $\sigma$ in the Bloch coherent state basis \cite{Livine:2007vk} is defined as
\begin{align}
Z=\sum_{\{j_f\}}\int dg_{ve}\int dn_{ef}\prod_f  P_f.
\label{amplitudejn}
\end{align}
The sum is over spins $j_f\in \mathbbm N$, and the integrals are over the $Spin(4)$ gauge variables and $SU(2)$ variables labeling the edges.\footnote{We dropped a face normalization factor which is usually taken as the dimension $d(j_f)=2j_f+1$ of the $SU(2)$-irreducible Hilbert space. Other normalizations are possible but irrelevant in the present analysis.} The face amplitude $P_f$ is given by
\begin{align}\label{Pf}
P_f=\text{tr}\,\vec\Pi_{e\in f} P^+_{ef}\otimes P^-_{ef}
\end{align}
where $\vec\Pi$ denotes the ordered product (according to the cycle of edges) and
\begin{align}\label{Pef}
P^\pm_{ef}=g^\pm_{e,s(e)}\varket{n_{ef}}^{\otimes2j^\pm_f}\bra{n_{ef}}^{\otimes2j^\pm_f}(g^\pm_{e,t(e)})^{-1}.
\end{align}
Here $\ket{n}$ is the $SU(2)$ Bloch coherent state \cite{Bloch:1946zza} for angular momentum\footnote{We introduced the notation $\varket{j,n}$ for the standard antilinear map $\epsilon$ applied to $\ket{j,n}$. In the standard basis, it is given by the symbol ${}^{j}\epsilon_{mm'}=(-1)^{j+m}\delta_{m,-m'}$.} along the direction of $\vec n$, in the fundamental spin $1/2$ representation and we have the following constraint\footnote{Because of equation \eqref{ratio} $\gamma$ is quantized to be rational. This restriction is not present in the Lorentzian version of the theory.} on the spin labels:
\begin{align}\label{ratio}
j^\pm=\frac{1\pm\gamma}{2}j.
\end{align}
We have split the $SO(4)$ variables $g$ into selfdual and antiselfdual rotations $(g^+,g^-)\in SU(2)\times SU(2)$. The partition function \eqref{amplitudejn} can be written in the form of a path integral for an action as
\begin{align}
Z&=\sum_{\{j_f\}}\int dg_{ve}\int dn_{ef} e^{S},\label{amplitudejnS}\\
S&=\sum_f S_f=\sum_f \ln P_f.
\label{action}
\end{align}
\section{The continuum $\gamma\rightarrow 0$ limit}
The continuum limit of the theory is defined as the infinite refinement of the 2-complex \cite{Rovelli:2010qx} (possibly undergoing a second order phase transition to a smooth spacetime). Differently, here we define a continuum limit that is suitable to describe gravitational physics with a truncation of the theory on a finite cellular structure of spacetime, namely with an approximation of the full theory. The spinfoam amplitudes associated to finite graphs can be viewed as effective amplitudes obtained after a coarse-graining procedure \cite{Markopoulou:2002ja,Bahr:2010cq} applied to the spinfoam infinite `lattice'. But since the truncated amplitudes are not fundamental, there is no reason to keep fixed the Immirzi parameter to its `bare' value $\gamma_0$. The possibility of a  renormalization of the Barbero-Immirzi parameter has been recently advocated in different contexts \cite{Daum:2010qt,Benedetti:2011nd}.

Here we explore the possibility of a running towards zero, $\gamma\rightarrow 0$, simultaneously with the large-spin regime of the theory. Thus we consider
\begin{align}\label{limitconsidered}
j\rightarrow\infty,\;\gamma\rightarrow 0,\;j\gamma=const.
\end{align}
and $\gamma j$ is the macroscopic physical area in Planck area units. Notice that the Immirzi parameter controls the spacetime discreteness. In particular it controls the area gap and the spacing between area eigenvalues, thus the limit \eqref{limitconsidered} is the continuum limit for the area operator. 
%%%
%%%%

In order to see the effect of \eqref{limitconsidered} on the partition function, let us restrict our attention to a 2-complex which is dual to a simplicial triangulation: vertices are dual to 4-simplices, and edges are dual to tetrahedra. The analysis parallels the one of \cite{Conrady:2008mk,Barrett:2009gg} at fixed, large spins. We are interested in making explicit the dependence on the Immirzi parameter, so let us decompose the action \eqref{action} using \eqref{ratio} in the following way
%\begin{align}
%S_f=i &\gamma j_f \im (P^+_f-P^-_f)+i  j_f \im (P^+_f+P^-_f)+\nonumber\\
%+&\gamma j_f \re (P^+_f-P^-_f)+  j_f \re (P^+_f+P^-_f)
%\end{align}
\begin{align}\label{actionf}
S_f=a_f (\ln \tilde P^+_f-\ln \tilde P^-_f)+\frac{1}{\gamma}a_f (\ln \tilde P^+_f+\ln \tilde P^-_f)
\end{align}
where we have defined the area of the triangle dual to $f$ as $a_f=\gamma j_f$, and $\tilde P^\pm_f$ is the face amplitude in the fundamental representation. We want to evaluate the partition function \eqref{amplitudejnS} in a region of macroscopic areas $a_f$ and in the limit $\gamma\rightarrow 0$ with $a_f$ fixed (so $j_f\rightarrow\infty$).  Collecting all the face terms \eqref{actionf}, let us write the full action as
\begin{equation}\label{fullaction}
S=S^{0}+\frac{1}{\gamma}S'.
\end{equation}
In the limit $\gamma\rightarrow 0$ with $a_f$ fixed the partition function can be approximated with an integral\footnote{We are only interested in the oscillatory behaviour of the integral, so we drop one global factor $1/\gamma$ per each face, that takes into account the measure of the Riemann sum.} over continuous areas
\begin{align}\label{integral}
Z\simeq\int da_f\int dg_{ve}\int dn_{ef} e^{S^0+\frac{1}{\gamma}S'}.
\end{align}
 For the stationary phase evaluation of \eqref{integral} we have to take variations of the second term $S'$ proportional to the large parameter $\frac{1}{\gamma}\rightarrow\infty$. The action $S'$ is complex with nonpositive real part $\text{Re}\,S'\leq 0$, so the main contribution to the integral comes from the critical points, where $\text{Re}\,S'=0$. This condition holds for $\text{Re}\,P_f^+=\text{Re}\,P_f^-=0$. One can easily show that the critical points are the solutions to
\begin{align}
g^+_{ev}\vec n_{ef}=-g^+_{e'v}\vec n_{e'f}\label{critical+}\\
g^-_{ev}\vec n_{ef}=-g^-_{e'v}\vec n_{e'f}
\label{critical-}
\end{align}
where $e,e'$ are adjacent edges in the face $f$, sharing the vertex $v$. If there are no solutions, the amplitude is exponentially suppressed. Now using \eqref{critical+}, \eqref{critical-} we have that the brackets
\begin{align}\label{brackets}
\bra{n_{ef}}(g_{ve}^{\pm})^{-1}g^\pm_{ve'}\varket{n_{e'f}}=e^{i\theta^\pm_{vf}}
\end{align}
reduce to simple phases, on the critical points (on-shell).

Furthermore, we must require that the critical points are also stationary. Varying $S'$ with respect to $SO(4)$ group variables, and evaluating at a critical point we get the condition
\begin{align}
\left.\delta_{g_{ev}} S'\right|_{crit.}=0\;\longrightarrow\;\sum_{f\in e}a_{f}\vec n_{ef}=0
\end{align}
which expresses the closure relation for the tetrahedron dual to the edge $e$. Variation with respect to the unit vectors $n_{ef}$ does not give further information, because it is automatically satisfied. Finally, we have to take variations of $S'$ with respect to the areas $a_f$, but this does not give further restrictions since
\begin{align}
\frac{\partial S'}{\partial a_f}=0
\end{align}
is automatically satisfied on the critical points. Indeed using \eqref{brackets} one can show that
\begin{align}
\left.\frac{\partial S'}{\partial a_f}\right|_{crit.}=\sum_{v\in f}(\theta^+_{vf}+\theta^-_{vf})=\Theta^*_f
\end{align}
is an angle with the interpretation of a torsion degree of freedom. But the on-shell discrete connection $g^\pm_{ev}$ is torsion-free (it is the discrete spin-connection) and this angle vanishes (see also \cite{Conrady:2008mk} for more details on torsion).

The critical points are in one-to-one correspondence with 4-dimensional spacetime triangulations (Regge manifolds), where the areas\footnote{It is well known that a generic assignment of areas does not correspond to any Regge triangulation \cite{Barrett:1997tx,Makela:2000ej,Dittrich:2008va}. In other words, in general there exists no assignment of lengths $l_s$ for the sides $s$ of triangles such that $a_f=a_f(l_s)$. The critical equations select only the Regge-like area configurations.} of the triangles are specified by $a_f$, and their 3d normals by $n_{ef}$. The Regge manifolds are endowed with a continuous, piecewise flat metric where curvature is distributional and concentrated on triangles. From a given critical point, one can reconstruct the area bivectors $A_{vf}$ of the triangles dual to $f$, in the frame of the 4-simplices $v$, namely the metric of the Regge manifold. 

In the $\gamma\rightarrow 0$ expansion the action must be evaluated at a critical point, and using \eqref{brackets} we have
\begin{align}
S|_{crit.}=i S_R=i \sum_f a_f \Theta_f
\label{Sfonshell}
\end{align}
where we have defined the deficit angle as
\begin{align}
&\Theta_f=\sum_{v\in f}(\theta^+_{vf}-\theta^-_{vf}).
\end{align}
The deficit angle $\Theta_f$ is the spacetime curvature concentrated on the triangle dual to the face $f$. The equation \eqref{Sfonshell} is the Regge form \cite{Regge:1961px,Williams:1991cd} of the action $S_R$ for general relativity. The asymptotic approximation of the integral \eqref{integral} requires to sum over all critical points. In general this will take the form of a continuous sum over the set $\mathcal C$ of critical points (critical manifold), with some measure $\mu$ that can be computed by standard asymptotic analysis tools. The result is
\begin{align}\label{previousintegral}
Z= \int_{\mathcal C} d\mu(a_f,n_{ef},g_{ev}) e^{i S_R}+\mathcal O(\gamma)
\end{align}
where $\mathcal O(\gamma)$ denotes the $\gamma$-corrections to the partition function coming from the next orders of the asymptotic approximation. 
%%
%%%
\section{The semiclassical $\hbar\rightarrow 0$ limit}
More interestingly, let us parametrize the previous integration \eqref{previousintegral} using length variables. Given that the critical points $(a_f,n_{ef},g_{ev})$ correspond to Regge triangulations, there exists an assignment of lengths $l_s$ to the sides $s$ of the triangles such that the areas $a_f$ coincide with the areas computed out of the lengths: $a_f=a_f(l_s)$.
Then we can parametrize the critical manifold $\mathcal{C}$ with the set of side lengths $l_s$. Restoring the $\hbar$ dependence, let us rewrite \eqref{previousintegral} as
\begin{align}\label{intlengths}
Z= \int d\tilde\mu(l_s) e^{\frac{i}{\hbar} S_R(l_s)}+\mathcal O(\gamma),
\end{align}
and the Regge action is now explicitly a function of the lengths
\begin{align}\label{Reggelengths}
S_{R}(l_s)=\sum_f a_f(l_s)\Theta_f(l_s)
\end{align}
as in the original classical formulation \cite{Regge:1961px}. The last expressions are a good starting point for taking the semiclassical limit. Before discussing this, let us make a few comments on the effect of sending the Immirzi parameter to zero. The remarkable consequence of the continuum limit $\gamma\rightarrow 0$ we have performed in the last section is that the spinfoam amplitude reduces effectively, that is up to $\gamma$-corrections, to a quantization of Regge gravity \cite{Rocek:1982fr} given by formula \eqref{intlengths} where the fundamental variables are continuous lengths. The result resonates with the recent findings in the computation of the graviton propagator within loop quantum gravity \cite{Bianchi:2006uf,Alesci:2007tx,Bianchi:2009ri}. In particular, as shown in \cite{Bianchi:2009ri}, the leading order (in the $\hbar$ expansion) graviton propagator $G(x,y)$ presents the same kind of $\gamma$-corrections,
\begin{align}\label{propagator}
G(x,y)= \frac{R+\gamma X+\gamma^2 Y}{|x-y|^2}+\hbar\text{-corr.}
\end{align}
and only in the limit $\gamma\rightarrow 0$ the tensorial structure of the 2-point function matches with the matrix of correlations $R$ computed in quantum Regge gravity, and, even more interestingly, with the one given by standard perturbative gravity on flat space. In retrospect, at the light of the present general analysis the previous result \eqref{propagator} is much more clear. Moreover, similar $\gamma$-corrections were found by Bojowald in the cosmological context \cite{Bojowald:2001ep}.

Suppose now we are interested in the semiclassical expansion of the spinfoam amplitude \eqref{intlengths}. This corresponds to looking at areas which are macroscopic, that is large as compared to the Planck area,
\begin{align}
\frac{a_f}{l^2_P}\gg 1,
\end{align}
or equivalently to the standard WKB expansion $\hbar\rightarrow 0$. This regime can be selected by appropriate semiclassical boundary conditions in the transition amplitudes, as explained in the next section.

The classical equations of  motion are obtained by varying  \eqref{Reggelengths} with respect to the lengths. Using also the Schlafli identity \cite{Regge:1961px}, which tells that the variation of the deficit angles do not contribute to the total variation of the action, these are the well-known Regge equations
\begin{align}\label{Ricciflat}
\sum_{f}\frac{\partial a_f}{\partial l_s}\Theta_f=0,
\end{align}
a discrete version of the continuum Einstein equations in vacuum, $R_{\mu\nu}=0$, namely of the vanishing of the Ricci tensor. They give a relation between the deficit angles of different faces. The integral \eqref{intlengths} is dominated by its stationary `trajectories', namely by the sets of lengths $l_s$ which are a solution of the Regge equations \eqref{Ricciflat}. However, in order to pick up a single classical `trajectory' we need to specify appropriately the boundary conditions and pass to the transition amplitudes. The spinfoam boundary formalism for the transition amplitudes is briefly reviewed in the next section.
%%%%
%%%%
%%%%
%%%%
\section{General boundary amplitudes} Given a 2-complex with boundary, its boundary graph $\Gamma$ is an abstract oriented graph made of links $l$ (where the external faces end) and nodes $n$ (where the external edges end). The boundary graph inherits its labeling from the external faces and edges. The set formed by the boundary graph $\Gamma$, the spins $j_l$ associated to the links, and unit vectors $n_{nl}$ associated to the nodes is the boundary data.
The spinfoam transition amplitude for the 2-complex $\sigma$ with boundary $\Gamma$ in the Bloch coherent state basis \cite{Livine:2007vk} is a functional of the boundary data\footnote{The amplitude  \eqref{amplitudejnSS} for a 2-complex with boundary has the nice interpretation as transition amplitude associated to a spin-network supported on the boundary graph $\Gamma$.} defined as
\begin{align}\label{amplitudejnSS}
W(j_l,n_{ne})=\sum_{\{j_f\}}\int dg_{ve}\int dn_{ef}\prod_l e^{S_l}\prod_{f} e^{S_f}
\end{align}
where now the action is split into a boundary action plus a bulk action. This formula needs further clarifications. First, the boundary action $S_l$ contains the external face amplitudes $P_l$. For an external face, the formula for $P_l$ is the same as \eqref{Pf} except that for the edges ending on the boundary we have only `half' of \eqref{Pef}. Second, the summation is over the sole internal spins $j_f\in \mathbbm N$, and the integrals are over the $Spin(4)$ gauge variables and $SU(2)$ variables (the unit vectors) labeling the internal edges.

The semiclassical analysis of the previous section is straightforwardly generalized to a 2-complex with boundary, which allows to select the continuum/semiclassical regime in the following way. First, write the boundary functional $W(a_l,n_{ne})$ in terms of the areas. Then we look at the behavior when all the boundary areas $a_l$ are macroscopic, $a_l\gg l_P^2$, in the limit $\gamma\rightarrow 0$ with $a_l$ fixed.\footnote{We suppose also that the path integral is dominated by areas of the same order of magnitude of the boundary areas: $a_l\simeq a_f$. In other words, the macroscopic boundary state must enforce macroscopic areas in the bulk of the triangulation. This is the condition of validity of the WKB expansion that has to be checked case by case for the specific geometry chosen.}

This regime is formally the simultaneous continuum $\gamma\rightarrow 0$ and semiclassical limit $\hbar\rightarrow 0$. Now the boundary amplitude gets its dominant contribution from the solution to the Regge equations \eqref{Ricciflat} compatible with the specified boundary data (see \cite{Sorkin:1975ah,Barrett:1994ks} on the initial value problem), namely we have the oscillatory behaviour
\begin{align}\label{result}
W(a_l,n_{ne})\sim e^{iS_R(a_l,n_{ne})},\quad\quad\gamma\rightarrow 0,\quad\quad\hbar\rightarrow 0
\end{align}
where now $S_R$ is the Hamilton function, that is the action evaluated at the classical trajectory determined by the boundary data.\footnote{If the boundary data are not consistent with a boundary triangulation, the amplitude is exponentially suppressed.} The set of boundary data can be equivalently mapped to the set of lengths of the boundary triangulation. This set of lengths is a Dirichelet boundary condition for the Regge equations of motion to be used to determine the classical solution in the interior\footnote{In the formula \eqref{result} we have supposed there is only one solution to the equations of motion with the assigned boundary. In the case there are many (a typical example is a continuous set of flat triangulated spacetimes) we should sum over them.} and evaluate the Hamilton function in \eqref{result}. 

The result \eqref{result} is consistent with what we expected from a theory of quantum gravity, and is the concrete realization of the equation \eqref{whatweexpect} in the introduction. 
%%%%
\section{Conclusions and outlook}
It this letter we have discussed a proposal for the semiclassical limit of spinfoams truncated to an arbitrary, finite triangulation (2-complex), where most calculations are done. We find (equation \eqref{result}) that the transition amplitudes are proportional to the exponential of the Hamilton function of Regge-Einstein general relativity, as expected, up to $\hbar$-corrections and $\gamma$-corrections in the simultaneous semiclassical and continuum limit. The first corrections correspond to the standard WKB expansion of the path integral. The latter are new and are the effect of the discreteness of geometry, in the sense that the spectra of areas and volumes are discrete and the discreteness is controlled by the Immirzi parameter $\gamma$. The continuum limit we take is a looser concept of the `full' continuum limit defined as the phase transition to a smooth spacetime manifold. However, working with a fixed, finite triangulation the only way of taking a continuum limit is to look at the continuous spectrum limit of the fundamental geometric operators. We have done this by letting $\gamma$ run to zero, keeping fixed the macroscopic areas $a_f$ (this is still a tentative proposal and must be further investigated). Remarkably, as explained in the previous section, the result sheds new light on the previous calculation of the graviton propagator with the `new models' \cite{Bianchi:2010zs}, where the same kind of $\gamma$-corrections to the standard perturbative tree-level propagator has been found (one could speculate on potentially observable signatures of those pure LQG corrections).

It is also very interesting to notice that essentially the same continuum limit was considered by Bojowald \cite{Bojowald:2001ep} in the context of loop quantum cosmology. Quoting its abstract: ``\emph{standard quantum cosmology is shown to be the simultaneous limit $\gamma\rightarrow 0$, $j\rightarrow\infty$ of loop quantum cosmology}'', a strong analogy it is worth studying further.

We have disregarded other possible contributions to the amplitude (symmetry related spacetimes, vector geometries, degenerate geometries etc.) that could spoil the correct semiclassical behaviour. For example we expect another term in the leading asymptotics \eqref{result} of the transition amplitudes which corresponds to a spacetime with opposite orientation that would change the (oscillatory part of the) amplitude into the sum of two sign-reversed exponentials, namely
\begin{align}
W(a_l,n_{nl})\sim \cos(S_R(a_l,n_{nl}))
\end{align}
as in the asymptotic formula for the amplitude of a single vertex \cite{PonzanoRegge:1968,Barrett:1998gs,Barrett:2009mw,Barrett:2009gg}. In principle this does not pose a real problem: for example, see \cite{Bianchi:2008ae} on the fate of the sign-reversed exponentials in the bulk of the triangulation, and how a coherent boundary state peaked on the appropriate extrinsic curvature is able to select a single exponential (see also \cite{Bianchi:2010mw}). Finally, we have also disregarded the potential divergencies associated to bubbles in the foam for which a suitable regularization and renormalization scheme \cite{Perini:2008pd,Krajewski:2010yq,Geloun:2010vj,Rivasseau:2011xg} could be  required. More details, including the Lorentzian signature, can be found in a longer version \cite{Magliaro:2011dz} of this letter.

\acknowledgments
We are grateful to Martin Bojowald and Carlo Rovelli for discussions. This work was supported in part by the NSF grant PHY0854743, The George A. and Margaret M.~Downsbrough Endowment and the Eberly research funds of Penn State. E.M. gratefully acknowledges support from ``Fondazione Angelo della Riccia''.

\bibliography{biblioreggefromsf}

\begin{thebibliography}{10}
\expandafter\ifx\csname url\endcsname\relax\def\url#1{\texttt{#1}}\fi

\bibitem{Baez:1997zt}
\Name{Baez J.~C.} \REVIEW{Class. Quant. Grav. }{15}{1998}{1827}.
%%CITATION = GR-QC/9709052;%%

\bibitem{Reisenberger:2000fy}
\Name{Reisenberger M. \and Rovelli C.} \REVIEW{}{}{2000}{}.
%%CITATION = GR-QC/0002083;%%

\bibitem{Perez:2003vx}
\Name{Perez A.} \REVIEW{Class. Quant. Grav. }{20}{2003}{R43}.
%%CITATION = GR-QC/0301113;%%

\bibitem{Misner:1957wq}
\Name{Misner C.~W.} \REVIEW{Rev. Mod. Phys. }{29}{1957}{497}.
%%CITATION = RMPHA,29,497;%%

\bibitem{Hawking:1979zw}
\Name{Hawking S.~W.} \REVIEW{Nucl. Phys. B}{144}{1978}{349}.
%%CITATION = NUPHA,B144,349;%%

\bibitem{Engle:2007wy}
\Name{Engle J., Livine E., Pereira R. \and Rovelli C.} \REVIEW{Nucl. Phys.
  B}{799}{2008}{136}.
%%CITATION = 0711.0146;%%

\bibitem{Livine:2007ya}
\Name{Livine E.~R. \and Speziale S.} \REVIEW{Europhys. Lett.
  }{81}{2008}{50004}.
%%CITATION = 0708.1915;%%

\bibitem{Freidel:2007py}
\Name{Freidel L. \and Krasnov K.} \REVIEW{Class. Quant. Grav.
  }{25}{2008}{125018}.
%%CITATION = 0708.1595;%%

\bibitem{Rovelli:1994ge}
\Name{Rovelli C. \and Smolin L.} \REVIEW{Nucl.Phys. B}{442}{1995}{593}.

\bibitem{Ding:2010ye}
\Name{Ding Y. \and Rovelli C.} \REVIEW{Class. Quant. Grav. }{27}{2010}{205003}.
%%CITATION = 1006.1294;%%

\bibitem{Rovelli:2010qx}
\Name{Rovelli C. \and Smerlak M.} \REVIEW{}{}{2010}{}.

\bibitem{Rovelli:2010bf}
\Name{Rovelli C.} \REVIEW{}{}{2010}{}.

\bibitem{Rovelli:2010ed}
\Name{Rovelli C. \and Speziale S.} \REVIEW{}{}{2010}{}.

\bibitem{Bianchi:2010bn}
\Name{Bianchi E., Han M., Magliaro E., Perini C., Rovelli C. \and Wieland W.}
  \REVIEW{}{}{2010}{}.

\bibitem{Han:2011as}
\Name{Han M. \and Rovelli C.} \REVIEW{}{}{2011}{}.

\bibitem{Bianchi:2006uf}
\Name{Bianchi E., Modesto L., Rovelli C. \and Speziale S.} \REVIEW{Class.
  Quant. Grav. }{23}{2006}{6989}.
%%CITATION = GR-QC/0604044;%%

\bibitem{Alesci:2007tx}
\Name{Alesci E. \and Rovelli C.} \REVIEW{Phys. Rev. D}{76}{2007}{104012}.
%%CITATION = 0708.0883;%%

\bibitem{Bianchi:2009ri}
\Name{Bianchi E., Magliaro E. \and Perini C.} \REVIEW{Nucl. Phys.
  B}{822}{2009}{245}.
%%CITATION = 0905.4082;%%

\bibitem{Bianchi:2010zs}
\Name{Bianchi E., Rovelli C. \and Vidotto F.} \REVIEW{Phys. Rev.
  D}{82}{2010}{084035}.
%%CITATION = 1003.3483;%%

\bibitem{Magliaro:2011qm}
\Name{Magliaro E. \and Perini C.} \REVIEW{Class.Quant.Grav.
  }{28}{2011}{145028}.

\bibitem{Conrady:2008ea}
\Name{Conrady F. \and Freidel L.} \REVIEW{Class.Quant.Grav.
  }{25}{2008}{245010}.

\bibitem{Conrady:2008mk}
\Name{Conrady F. \and Freidel L.} \REVIEW{Phys. Rev. D}{78}{2008}{104023}.
%%CITATION = 0809.2280;%%

\bibitem{Livine:2007vk}
\Name{Livine E.~R. \and Speziale S.} \REVIEW{Phys. Rev. D}{76}{2007}{084028}.
%%CITATION = 0705.0674;%%

\bibitem{Bloch:1946zza}
\Name{Bloch F.} \REVIEW{Phys.Rev. }{70}{1946}{460}.

\bibitem{Markopoulou:2002ja}
\Name{Markopoulou F.} \REVIEW{Class.Quant.Grav. }{20}{2003}{777}.

\bibitem{Bahr:2010cq}
\Name{Bahr B., Dittrich B. \and He S.} \REVIEW{}{}{2010}{}.

\bibitem{Daum:2010qt}
\Name{Daum J.-E. \and Reuter M.} \REVIEW{}{}{2010}{}.

\bibitem{Benedetti:2011nd}
\Name{Benedetti D. \and Speziale S.} \REVIEW{}{}{2011}{}.

\bibitem{Barrett:2009gg}
\Name{Barrett J.~W., Dowdall R.~J., Fairbairn W.~J., Gomes H. \and Hellmann F.}
  \REVIEW{J. Math. Phys. }{50}{2009}{112504}.
%%CITATION = 0902.1170;%%

\bibitem{Barrett:1997tx}
\Name{Barrett J.~W., Rocek M. \and Williams R.~M.} \REVIEW{Class.Quant.Grav.
  }{16}{1999}{1373}.

\bibitem{Makela:2000ej}
\Name{Makela J. \and Williams R.~M.} \REVIEW{Class.Quant.Grav.
  }{18}{2001}{L43}.

\bibitem{Dittrich:2008va}
\Name{Dittrich B. \and Speziale S.} \REVIEW{New J.Phys. }{10}{2008}{083006}.

\bibitem{Regge:1961px}
\Name{Regge T.} \REVIEW{Nuovo Cim. }{19}{1961}{558}.
%%CITATION = NUCIA,19,558;%%

\bibitem{Williams:1991cd}
\Name{Williams R.~M. \and Tuckey P.~A.} \REVIEW{Class.Quant.Grav.
  }{9}{1992}{1409}.

\bibitem{Rocek:1982fr}
\Name{Rocek M. \and Williams R.~M.} \REVIEW{Phys.Lett. B}{104}{1981}{31}.

\bibitem{Bojowald:2001ep}
\Name{Bojowald M.} \REVIEW{Class. Quant. Grav. }{18}{2001}{L109}.
%%CITATION = GR-QC/0105113;%%

\bibitem{Sorkin:1975ah}
\Name{Sorkin R.} \REVIEW{Phys.Rev. D}{12}{1975}{385}.

\bibitem{Barrett:1994ks}
\Name{Barrett J.~W., Galassi M., Miller W.~A., Sorkin R.~D., Tuckey P.~A.
  \etal} \REVIEW{Int.J.Theor.Phys. }{36}{1997}{815}.

\bibitem{PonzanoRegge:1968}
\Name{Ponzano G. \and Regge T.} \REVIEW{}{}{}{} spectroscopic and Group
  Theoretical Methods in Physics, edited by F.Block (North Holland, Amsterdam,
  1968).

\bibitem{Barrett:1998gs}
\Name{Barrett J.~W. \and Williams R.~M.} \REVIEW{Adv.Theor.Math.Phys.
  }{3}{1999}{209}.

\bibitem{Barrett:2009mw}
\Name{Barrett J.~W., Dowdall R.~J., Fairbairn W.~J., Hellmann F. \and Pereira
  R.} \REVIEW{Class. Quant. Grav. }{27}{2010}{165009}.
%%CITATION = 0907.2440;%%

\bibitem{Bianchi:2008ae}
\Name{Bianchi E. \and Satz A.} \REVIEW{Nucl.Phys. B}{808}{2009}{546}.

\bibitem{Bianchi:2010mw}
\Name{Bianchi E., Magliaro E. \and Perini C.} \REVIEW{Phys.Rev.
  D}{82}{2010}{124031}.

\bibitem{Perini:2008pd}
\Name{Perini C., Rovelli C. \and Speziale S.} \REVIEW{Phys. Lett.
  B}{682}{2009}{78}.
%%CITATION = 0810.1714;%%

\bibitem{Krajewski:2010yq}
\Name{Krajewski T., Magnen J., Rivasseau V., Tanasa A. \and Vitale P.}
  \REVIEW{Phys.Rev. D}{82}{2010}{124069}.

\bibitem{Geloun:2010vj}
\Name{Geloun J.~B., Gurau R. \and Rivasseau V.} \REVIEW{Europhys.Lett.
  }{92}{2010}{60008}.

\bibitem{Rivasseau:2011xg}
\Name{Rivasseau V.} \REVIEW{}{}{2011}{}.

\bibitem{Magliaro:2011dz}
\Name{Magliaro E. \and Perini C.} \REVIEW{}{}{2011}{} * Temporary entry *.

\end{thebibliography}
\bibliographystyle{eplbib}
\end{document}